# Possibility of superconducting gap increase in heterogeneous superconductors

*I. N. Zhilyaev*

*Institute of Microelectronics Technology and High Purity Materials, Russian Academy of Sciences,*

*Chernogolovka, Moscow Region, Russia*

zhilyaev@ipmt-hpm.ac.ru

*A picture of pair scattering on Wigner crystal islands for the earlier proposed superconducting state in BCS superconductor with characteristic HTSC properties is presented. The model is supposed to be valid for thin heterogeneous films and boundaries between superconductors.*

In work [1] we put forward an idea of a superconducting state for a heterogeneous quasi-two-dimensional superconductor having a structure of a homogeneous matrix with island-like impregnations of the lateral size d about ten Angstroms and distances between the islands of the same order. The matrix has the properties of a standard BCS superconductor, and the islands possess both the properties of a Wigner charging density wave and a BCS superconductor, depending on the charge density. The characteristic features of the proposed model resemble the properties of HTSC. In some sense, the idea of the model is related to those proposed by Little for a quasi-one-dimensional case and by Ginzburg for a quasi-two-dimensional case [2]. Basically, their idea was to look for a pairing interaction of a higher energy in an electron system than that of electron-phonon interaction and, thus, to enhance superconducting characteristics. The idea of our model is also to find a large energy in an electron system. However, in our case the superconducting state is formed in a different way: the pairing potential is determined in accordance with the standard BCS model at electron-phonon interaction, but the probability of electron states filling, which in the BCS model is formed by a self-consistent procedure from electron-phonon interactions, is determined in our case by the interaction of electron pair states with CDW islands. In other words, the proposed model suggests including pair interaction through CDW islands in addition to electron-phonon interaction.

In [1], pair interaction with quantum fluctuations of CDW islands was considered as additional interaction. However, such interaction is questionable, and, if realized, its effect on the superconducting state could be destructive, similar to the influence of thermal fluctuations. In this work we propose another way of pair interaction with CDW islands which is similar to pair electron interaction through phonons. But unlike the interaction through phonons which changes the lattice state and provides both the pairing energy and energy for pair state smearing, this model suggests pair interaction of matrix electrons through island electron states, which modifies the state of islands. As a consequence, the arising additional smearing of pair electron states enhances the superconducting state.

The universal dependence of critical superconducting transition temperature on the density of charge carriers, low density of charge carriers, and quasi-two-dimensional character of conductivity in HTSC cuprates suggests the involvement of Wigner type CDW in the mechanism of superconductivity in cuprates. Let Wigner type CDW islands with the $C_4$ symmetry and lateral sizes close to the CDW wave length be present in a quasi-two-dimensional superconductor. Also, let the redistribution of charge in islands be spatially symmetric in both directions and one CDW period correspond the d size (see fig.). Assume that the concentration of charge is such that the CDW state of islands is close to the phase transition to a state without CDW at an increase of the concentration of charge carriers in islands. The state without CDW at an increase of the concentration is, therewith, metallic enabling the BCS pairing through phonons. In the Wigner case the number of charge carriers in an island is close to unity. Then, a change in the island charge by a value of unity can result in a radical change in the island energy state. In our case, the effective number of charge carriers in the island can change during the matrix charge carrier scattering. At an increase of the island charge, this would lead to the transition of the island to a state without CDW. Under the conditions when the time of charge carrier interaction with an island $\tau_d$ is close to the time of energy change of the island state, the interaction of charge carriers with the island could be inelastic. Owing to the spatial symmetry of island charge distribution and spatial symmetry of redistribution processes, only pair processes of scattering are allowed. Usually the probability of pair processes is small. But in HTSC the Fermi energy is small, Fermi impulses are also small, and the corresponding wave lengths of charge carriers can be of several Angstroms, i.e. about the island sizes. In such conditions the probability of pair processes can be of the order of unity. As was mentioned above, the efficiency of a scattered electron interaction with an island depends on the ratio of interaction time $\tau_d = d/v_F$ to the time of energy state changes of the island. The value of $\tau_d \cong$ $10^{-14}$ sec at the characteristic HTSC Fermi velocity $v_F \cong 10^7$ cm/sec. Let estimate the time of changes in the island state charge. The island transitions between the states with and without CDW give rise to spatial redistribution of electron density and Coulomb interaction which should result in the plasma law of oscillations. Plasma waves in a two-dimensional conductor are described by the law of dispersion $1/\omega^2 = 1/(c/\lambda)^2 + 2/\omega_p^2$ [3], where c is the velocity of light and $\omega_p = (4\pi n_e e^2/m)^{1/2}$ is the plasma frequency of a three-dimensional conductor (e and m are the charge and weight of an electron). The plasma frequency for charge density in HTSC near an optimum doping $n_e \cong 10^{21}$ cm$^{-3}$ is estimated to be $\omega_p \cong 1.8 \cdot 10^{15}$ radian/sec. In our case the wave length $\lambda$ cannot be larger than the size d. Then oscillations can have only one frequency $\omega_s = \omega_p/\sqrt{2} = 1.3 \cdot 10^{15}$ radian/sec. The time characterizing the dynamics of island transition from the state with to state without CDW $\tau_s = 2\pi/\omega_s \cong 0.5 \cdot 10^{-14}$ sec is comparable with $\tau_d$. Thus, in our case the scattering processes can be inelastic. As a result of pair inelastic processes of scattering by islands with CDW, the distribution of pair states by energy can be smeared by a value of the order of the CDW gap.

Numerical estimations of superconducting parameters for the given picture of scattering coincide with those in [1]. Accepting the observed large value of a pseudogap $\varepsilon_{pg}$ as an CDW gap in HTSC and providing the Debye energy $\hbar\omega_D$ is somewhat larger than the pseudogap $\varepsilon_{pg}$, the value of the probability of electron states filling $v_k^2$ can be close to 1/2 in the range of the pseudogap values, i.e. in a much greater range than in the usual BCS model. Thus, unlike the standard BCS model, these are the changes in the state of CDW island electron system that provide the energy to cause $v_k^2$ smearing rather than changes in the phonon system state. If this smearing is taken into account, the superconducting gap $\Delta$ radically increases even in the case of weak coupling at electron-phonon interaction and for HTSC parameters gives a value close to that observed for HTSC [1]. The effect of superconductivity enhancement would decrease at an increase or a decrease of the charge carrier concentrations because these concentrations would not correspond to the Wigner crystal conditions.

Now consider the pair scattering on CDW islands in more detail. Let Coulomb repulsive energy of charge carriers, connected with CDW of Wigner type, transform into kinetic energy because of the island charge increase at the island-pair interaction. Note that no attraction potential exists in this interaction as opposed to electron pair interaction through phonons. At the same time, as the island charge is large enough during scattering, its state corresponds to that of metal where coupling through phonons is possible. Because pair interaction with islands occurs virtually in a non-interruption manner, i.e. all islands are involved in scattering, they should not hamper coupling through phonons. Note also that spatial localization of CDW islands should not interfere with delocalized processes of coupling through phonons. The matter is that because of low Fermi energy and, hence, large wave lengths corresponding to charge carriers, the interaction of pairs with CDW islands can be considered delocalized at a distance of about coherence length (several Angstroms).

One more feature of the proposed scattering picture is worth noting. At the border of CDW islands and matrix there should be a potential barrier. Its value can be of the order of the CDW gap (for HTSC it is probably about $\varepsilon_{pg}$). Pairs should have energy of the order $\varepsilon_{pg}$ to overcome the barrier in the case of inelastic interaction with islands. This could be the case if $\hbar\omega_D > \varepsilon_{pg}$. Under such conditions the energy range with $v_k^2$ close to 1/2 does not depend on $\hbar\omega_D$, therefore the isotopic effect can be close to zero, and at an increase and a reduction of charge carrier concentration it returns to the value characteristic for BCS like in the case of HTSC. In other words, the proposed picture of scattering on CDW islands suggests that the isotopic effect is close to zero in the maximum of $\Delta(n_e)$ dependence.

The model can probably explain an observed essential increase of the critical temperature $T_C$ of superconducting transition in heterogeneous thin films in some instances as compared with initial critical temperatures of bulk superconductors. The number of proposed mechanisms available now [4] shows that the situation in this area has not been clarified yet. In granulated films in immediate proximity (about a 100-th fraction) to the temperature of transition to the superconducting state, an observed rise in the temperature of superconducting transition in some materials can be explained in terms of the Aslamazov-Larkin paraconductivity theory [5] for a two-dimensional case. However, large effects in heterogeneous films are hard to explain on the basis of this theory [4]. It is interesting to note that the greatest effects in heterogeneous films are observed as the granule sizes approach the range of several Angstroms [4], which is close to heterogeneity scale in HTSC [6]. Then it seems pertinent to use the proposed mechanism to explain such effects in thin films. Besides, it is known from the literature, that CDW are usually present in layered quasi-two-dimensional materials. Assume that thin superconducting films are also two-dimensional heterogeneous objects and CDW islands can arise in them. Then, according to the proposed mechanism the effect of enhancement of superconducting properties can be observed in them. It is of interest to estimate the enhancement effect and to compare it with values observed experimentally. Certainly, such estimation is difficult to make because this requires knowledge of island distribution in the sample and parameters of CDW, but it is possible to compare the orders of magnitudes as it was made for HTSC in [1]. When the probability of filling of electron states $v_k^2$ is given, $\Delta$ at the temperature close to zero can be estimated with the help of expression [1]

$$\Delta = V \sum_k v_k \sqrt{(1 - v_k^2)}$$

Here V is the electron interaction potential. Let substitute integration on energy $\varepsilon$ for the summation on states. To estimate the order of magnitude, let use the formula for an isotropic case, then

$$\Delta \cong V \int_{-\delta\varepsilon_{hc}/2}^{\delta\varepsilon_{hc}/2} n(E_F)(1/2)d\varepsilon$$

where $n(E_F)$ is the density of states at the Fermi level and $\delta\varepsilon_{hc}$ is the boundary of smearing energy to which $v_k^2 \approx 1/2$. In this case assume that $\delta\varepsilon_{hc}$ can be of the order of the Debye temperature. Let estimate $T_c$ for aluminum where the observed effect is rather great. Value $n(E_F)V \approx 0.2$, and the Debye temperature is 400K for aluminum. Then, considering the connection between $T_c$ and $\Delta$ to be the same as in BCS, find $T_c \approx 40K$, which is by an order of magnitude larger than the observed value [4]. Because in real structures such parameters as CDW gap of islands and efficiency of pair interaction with islands can be less than those accepted for estimation, then the fact that an estimated maximum value exceeds the observed ones is indicative of a large potentiality of the proposed model to explain such effects.

Probably, the proposed mechanism can explain the observed increase of critical temperature in the area of a quasi-two-dimensional boundary of two superconductors [7].

The author is grateful to S.A.Brazovskiy, V.M.Edel'shtein, I.A.Larkin, L.N.Zherihina, M.Yu.Barabanenkov for the consultations and useful discussion.

This work was supported by the program "Quantum mesoscopic and disordered structures" of the Russian Academy of Sciences.

1. I.N.Zhilyaev, Low Temperature Physics, **38**, 1063 (2012).
2. Eds V.L.Ginzburg, D.A.Kirzhnits. High-Temperature Superconductivity. New York, Consultant Bureau, 1982.
3. T.Ando, A.Fauler, F.Stern, Reviews of Modern Physics, **54**, (1982).
4. Yu.F.Komnik. Physics of Metal Films. Moscow, Atomizdat, 1979.
5. L.G.Aslamasov, A.I. Larkin, Phys.Lett. A, **v.26**, 238 (1968).
6. W.D. Wise, M.C. Boyer, et.al. Nature Phys., **v.4**, 696 (2008).
7. M.S. Khaikin, I.N. Khlyustikov, JETP.Lett., **v.33**, 158 (1981).

Fig. A schematic illustration of charge redistribution in one of the two principal directions. The lines in the figure represent the extremes of a charge density wave with the period d. The arrows show the directions of the charge redistribution.

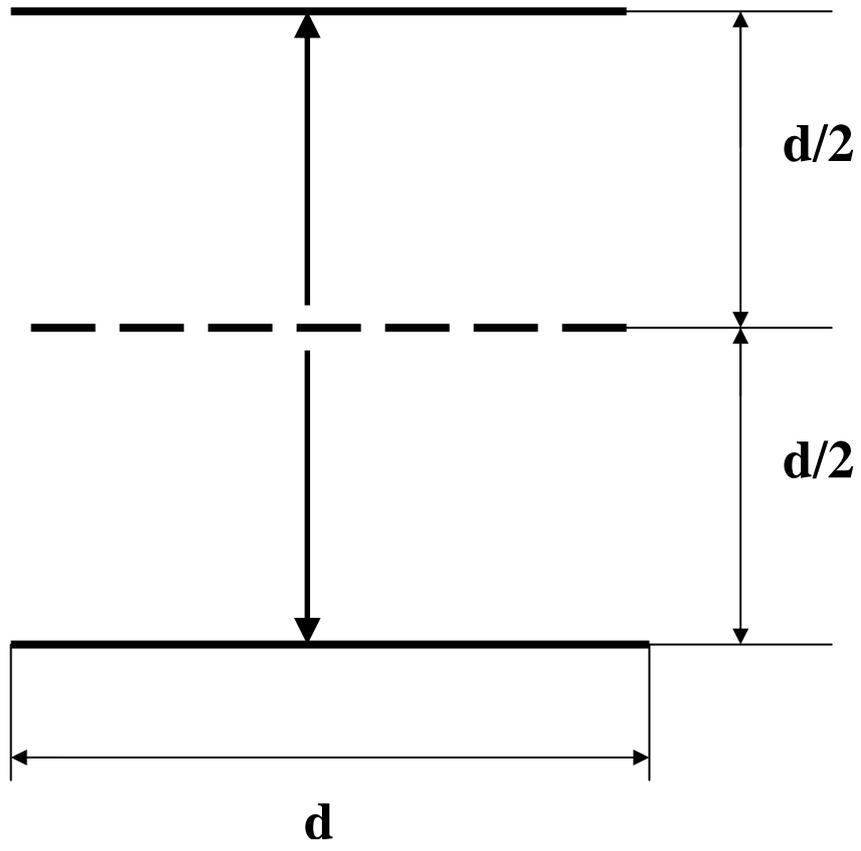